\documentclass[12pt]{iopart}
\expandafter\let\csname equation*\endcsname\relax
\expandafter\let\csname endequation*\endcsname\relax
\usepackage{amssymb}
\usepackage{graphicx}
\usepackage{physics}
\usepackage{xcolor}

\def\TT{\textcolor{black}}
\def\RM{\textcolor{black}}

\begin{document}

\title[Breakdown of quantization in nonlinear Thouless pumping]{Breakdown of Quantization in nonlinear Thouless pumping}

\author{T Tuloup$^1$, R W Bomantara$^2$ and J Gong$^{1,3,4,5}$}
\address{$^1$ Department of Physics, National University of Singapore, Singapore 117543}
\address{$^2$ Department of Physics, Interdisciplinary Research Center for Intelligent Secure Systems, King Fahd University of Petroleum and Minerals, 31261 Dhahran, Saudi Arabia}
\address{$^3$ Center for Quantum Technologies, National University of Singapore, Singapore
117543, Singapore}
\address{$^4$ Joint School of National University of Singapore and Tianjin University,
International Campus of Tianjin University, Binhai New City, Fuzhou 350207, China}
\address{$^5$ MajuLab, CNRS-UCA-SU-NUS-NTU International Joint Research Unit, Singapore}
\eads{\mailto{Raditya.Bomantara@kfupm.edu.sa}, \mailto{phygj@nus.edu.sg}}

\begin{abstract}
The dynamics of solitons driven in a nonlinear Thouless pump and its connection with the system's topology were recently explored for both weak and strong nonlinear strength. \TT{Using both a self-consistent algorithm and 4th order Runge Kutta method,} this work uncovers the fate of nonlinear Thouless pumping in the regime of intermediate nonlinearity, thus establishing a fascinating crossover from the observation of nonzero and quantized pumping at weak nonlinearity to zero pumping at strong nonlinearity. We identify the presence of critical nonlinearity strength at which quantized pumping of solitons breaks down regardless of the protocol time scale. Such an obstruction to pumping quantization is attributed to the presence of self-crossing in nonlinear topological bands. By considering another type of pumping involving Bloch states, we further show how the presence of self-crossing bands also leads to breakdown of quantization, but in a completely different manner from that in the case of soliton pumping. Our results not only unveil a missing piece of physics in nonlinear Thouless pumping, but also provide a means to detect loop structures of nonlinear systems investigated in real space and momentum space.
\end{abstract}
\vspace{2pc}


\submitto{\NJP}
\maketitle
%
%
%
%
%

\section{Introduction}

Thouless pumping is one of the most prominent examples of quantized topological transport~\cite{Thouless1983,Niu_1984,ShortCourseTopologicalInsulators}. Since its theoretical prediction by Thouless, it has been experimentally realized in many systems such as ultracold atom systems, both bosonic~\cite{Lohse2016ThoulessBosonic,Lohse2018ThoulessBosonic,Nakajima2021ThoulessBosonic} and fermionic~\cite{Nakajima2016ThoulessFermionic}, photonics~\cite{Kraus2012ThoulessPhotonics,Verbin2015ThoulessPhotonics,Zilberberg2018,Cerjan2020ThoulessPhotonics}, or spin systems~\cite{Ma2018ThoulessSpin}. Among these systems, many are playgrounds for interesting nonlinear effects such as the Kerr effect naturally arising in optical lattices~\cite{SelfLocStatesPhotonicTI,TopologicalNatureNLOpticalEffectsSolids,TopoTransitioninSSHthroughNLKerr,KerrEffectNonlinearTopologicalPhotonics,NonlinearTopologicalPhotonics}, or the many-body interacting behavior of Bose-Einstein condensates of bosonic cold atoms~\cite{DarkSolitonsBEC,GenerateSolitonsPhaseEngiBEC,MatterWaveSolitonTrains,Review_BEC_2003,ExampleMeanFieldBEC,ReviewUltracoldAtoms}. 

The behaviour of the above-mentioned nonlinear systems is best described by the nonlinear Schr\"{o}dinger equation known as the Gross-Pitaevskii (GP) equation, where the nonlinear terms come from the mean-field treatment of the interacting bosons~\cite{GrossBEC_GP,PitaevskiiBEC_GP}. This type of nonlinear problem has been extensively studied in recent years, and has been shown to exhibit several exotic features, two of which are especially relevant for the present work. First, it is known that the energy spectrum of some nonlinear systems (to be referred to as nonlinear energy spectrum hereafter) with respect to some parameter may develop self-crossing bands such as looped-bands or swallowtail structures~\cite{NL_LZ_Tunneling,Diakonov2002Loop,Machholm2003Loop,Review_BEC_2003,Morsch2006Loop,ReviewUltracoldAtoms,Zhang2008,Eckel2014Loop,Lim2020}. Second, certain nonlinear systems are known to support solitons~\cite{Chiao1964Soliton,Askar_yan_1974Soliton,ablowitz1981solitons,Christodoulides1988Soliton,Eisenberg1998Soliton,Stegeman1999Soliton,Lumer2013Soliton,Leykam2016EdgeSoliton,ChiralityGapSoliton,Mulherjee2020FloquetSoliton}. In this manuscript, we refer to solitons as localized wavepackets that arise at finite nonlinearity.

In exploring physics arising from the interplay of nonlinearity and topology, the Thouless pumping of such solitons has been experimentally observed~\cite{Jurgensen2021ExperimentalSolitonPumping}, and theoretically studied in both the weak nonlinear regime~\cite{Jurgensen2022,mostaan2022quantized,Fu2022} and the strong nonlinear regime~\cite{Fu2022,Jurgensen2023}. In the fermionic case, the breakdown of quantized Thouless pumping in the presence of Hubbard interactions has been experimentally demonstrated~\cite{Walter2022BreakdownThoulessPump}. Among more recent developments, nonlinearity is known to fundamentally modify the system’s adiabatic following dynamics, leading to path-dependent dynamical phases that modify the Berry connection and, consequently, the Aharanov-Bohm phase~\cite{NLDC} and the Zak phase~\cite{NLZP}. This motivated us to reexamine the nonlinear Thouless pumping especially if its quantization sustains due to nonlinearity.

Using a typical model system plus nonlinearity, here we demonstrate that a dynamically evolved state during a pumping protocol may differ from the instantaneous nonlinear stationary state no matter how slow the protocol is, even if the corresponding linear model remains gapped throughout the whole process. This puzzle is explained by finding the presence of self-crossing bands (to be further elaborated below) in the energy spectrum as a function of some adiabatic parameter. The emergence of self-crossing bands in the energy spectrum at moderate nonlinearity marks the breakdown of adiabaticity, which then leads to non-quantized pumping. We hence discover a mechanism accounting for a definite breakdown of quantized soliton pumping, serving as a crossover from quantized nonzero pumping at weak nonlinearity~\cite{Jurgensen2022,mostaan2022quantized,Fu2022} to zero pumping at strong nonlinearity~\cite{Fu2022}. 

To gain another perspective on the role of self-crossing bands in altering the pumping dynamics, we consider a different but analytically simpler type of pumping involving Bloch states, then explicitly derive the nonlinear correction to the displacement that in turn also leads to the breakdown of quantization. In this case, we find that the critical nonlinearity strength at which loop structures start to appear corresponds to an extremum for the nonlinear drift and a singularity in both the Berry curvature and its nonlinear correction.

To gain analytical insight into the role of loop structures in altering the pumping dynamics, we explicitly derive the nonlinear correction in the particle displacement for a Bloch state. Specifically, we find that the critical nonlinearity strength at which loop structures start to appear corresponds to an extremum for the nonlinear drift and a singularity in both the Berry curvature and its nonlinear correction.

This paper is organized as follows. In section~\ref{section:Nonlinear Thouless pump}, we introduce our model of a nonlinear Thouless pump, and present the dynamical evolution of solitons for different regimes of nonlinearity strength. Major results include the failure of adiabatic following no matter how slow the protocol is, with the main peak of the soliton still being displaced in spite of some spreading all over the lattice. In section~\ref{section:Breakdown of the adiabatic due to the emergence of loop structures}, we study the nonlinear energy spectrum of our model. We highlight the presence of self-crossing bands for a certain range of nonlinearity, and show how it correlates with three very distinguishable behaviors for the pumped soliton. In section~\ref{section:Nonlinear adiabatic pumping in momentum space} we study the pumping of a Bloch state in our model and derive an additional contribution to the average displacement of a particle due to nonlinear dynamics. In section~\ref{section:Discrepancy between real space and momentum space pumping} we highlight the crucial differences between the two types of pumping considered, and discuss how self-crossing bands developed in nonlinear energy bands break the quantization of both pumpings in different ways. In section~\ref{section:Concluding remarks} we summarize the main findings of this work and suggest possibilities for future studies.

\section{Nonlinear Thouless pump}
\label{section:Nonlinear Thouless pump}

In this work, we consider a nonlinear Thouless pump, consisting of a chain of $N$ dimers. It is made up of a time dependent Rice-Mele model to which we add focusing nonlinearity \cite{RiceMeleModel}. The resulting system is described by the following set of nonlinear Schr\"{o}dinger equations for $j = 0 \, \ldots \, N-1 $,
\begin{eqnarray}
\label{eqn:NonlinearRealSpaceModel}
\rmi \frac{\rmd \Psi_{2j}}{\rmd t} = v(t) \Psi_{2j+1} + w(t) \Psi_{2j-1} + [u(t) - g \left|\Psi_{2j}\right|^2 ] \Psi_{2j}\\
\rmi \frac{\rmd \Psi_{2j+1}}{\rmd t} = v(t) \Psi_{2j} +  w(t) \Psi_{2j+2} - [u(t) + g \left|\Psi_{2j+1}\right|^2] \Psi_{2j+1}
\end{eqnarray}
where $v(t) = - [J + \delta \sin{\omega t}]$ is the intracell coupling, $w(t) = - [J - \delta \sin{\omega t}]$ is the intercell coupling, $u(t) = - \Delta \cos{\omega t}$ is the staggered on-site potential, $g > 0$ is the strength of the focusing Kerr-like nonlinearity, and $\omega$ is the modulation frequency, taken small enough to allow adiabatic evolution. Here, adiabaticity is defined in a similar manner as its linear counterpart. Namely, it refers to the condition under which an initially prepared stationary state evolves into another stationary state of a set of slowly varying nonlinear Schr\"{o}dinger equations. We further refer to the latter as adiabatic following. \TT{The parameters of the system are taken to be periodically driven \RM{sufficiently slowly}, so that in the linear limit $g=0$ the model realizes adiabatic charge pumping. In the linear case, the lower band has a Chern number of 2, meaning that a wavepacket uniformly occupying the lower band will be displaced by 2 sites over one adiabatic pumping cycle. \RM{The corresponding nonlinear model} is \RM{a type of} a time-dependent GP equation, and although it was historically used to describe the dynamics of Bose-Einstein condensates of ultracold bosonic matter~\cite{GrossBEC_GP,PitaevskiiBEC_GP}, it is equally useful to describe the propagation of pulsed light through an array of waveguides, the most popular experimental platform to realize nonlinear pumping of solitons~\cite{Jurgensen2021ExperimentalSolitonPumping}. In this case, the nonlinear term is caused by the optical Kerr effect.} For all calculations below, we use $J=1$, $\delta=0.5$ and $\Delta=1$, with $N=100$ unit cells ($2N = 200$ sites), and we impose open boundary conditions (OBC) with $\Psi_{-1} = \Psi_{2N} = 0$. All physical variables presented in this work are assumed to be scaled, and therefore are in dimensionless units.

In the following, we will compare the dynamics of a soliton throughout the pumping cycle obtained from two different methods. In the first method, the soliton at time $t$ is obtained by finding the stationary state of equation~(\ref{eqn:NonlinearRealSpaceModel}) via an iterative procedure detailed in \ref{app:A}, using the instantaneous soliton at the previous time step as the trial state. In the second method, the soliton at time $t$ is obtained by directly solving equation~(\ref{eqn:NonlinearRealSpaceModel}) using 4th order Runge Kutta method under a given initial soliton. \TT{Note that particular care must be taken in the implementation of the latter as the nonlinear term that can strongly vary even for small time steps. \RM{To this end,} we updated the rate of change in the method at each step based of the current state of the system} \RM{and additionally checked the convergence of the method by comparing solutions at different time steps}. The first method corresponds to a perfect adiabatic following, meaning that at any given time the soliton obtained is an instantaneous stationary state of the nonlinear Hamiltonian. On the other hand, the second method corresponds to the actual dynamical evolution of the soliton and depends on the modulation frequency $\omega$. In experiments, the soliton dynamics is then expected to follow the description of the second method. Under the same initial soliton and assumption of adiabaticity, it is na\"{i}vely expected that the soliton dynamics obtained from both methods coincide at all time. We will however see that this is not the case for some range of nonlinearity, hinting for a breakdown of the adiabatic process. In our numerical studies, we prepare our initial soliton to be localized around site $n=100$, which is found using an iterative, self-consistent algorithm presented in \ref{app:A} under a trial hyperbolic secant profile wavepacket.

\begin{figure}
  \centering
  \includegraphics[width=0.8\columnwidth]{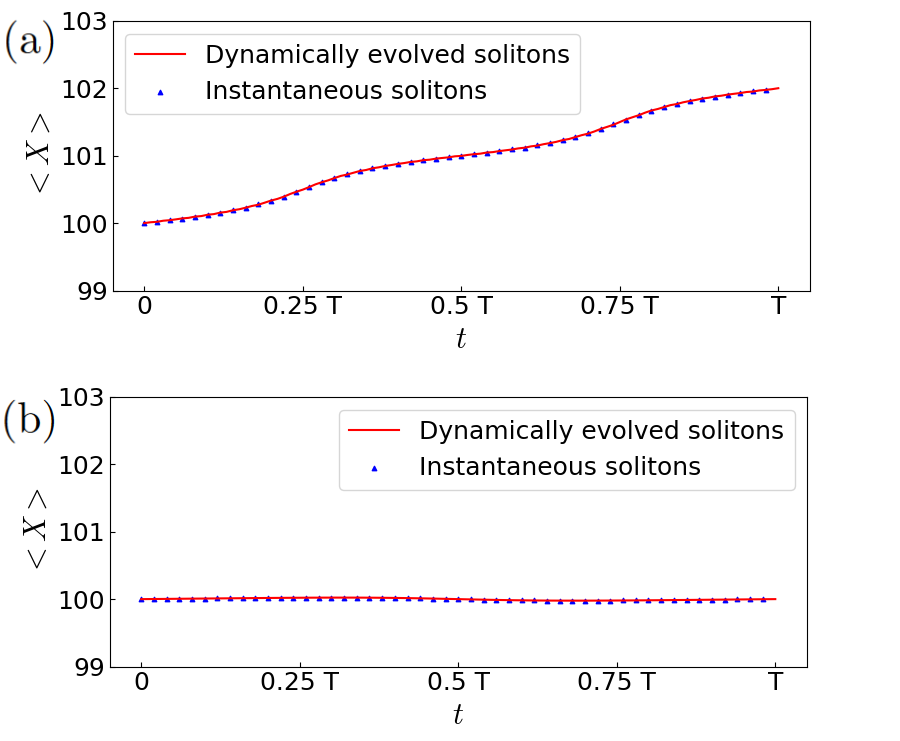}
  \caption{Comparison of the expectation value of position between instantaneous and dynamically evolved solitons over a pumping cycle, for different nonlinearity strength of a) $g=1$ and b) $g=10$. The modulation frequency was taken as $\omega = 10^{-3}$ in all panels.}
  \label{fig:AdiabaticFollowingWeakStrong}
\end{figure}

In figure~\ref{fig:AdiabaticFollowingWeakStrong}, we show the position expectation value $\langle X \rangle = \sum_j j \left|\Psi_{j}\right|^2 $ of the soliton, hereafter to be equivalently referred to as the position of the center of mass (COM) of the soliton, over one adiabatic cycle following the two methods above. Figure~\ref{fig:AdiabaticFollowingWeakStrong}-a) shows that for weak nonlinear strength, both methods yield identical results at all time; the evolution of the soliton remains adiabatic, and the total displacement over one cycle is quantized to the Chern number of a band in the linear limit. This quantized displacement was explained in \cite{Jurgensen2022, mostaan2022quantized,Fu2022} in terms of the Wannier states expansion of the soliton. On the other hand, in the strongly nonlinear regime, shown in figure~\ref{fig:AdiabaticFollowingWeakStrong}-c), the soliton merely oscillates around the same site and yields zero displacement over one cycle. This absence of pumping can be explained by the nonlinearity induced Rabi oscillations between two of the lowest bands, whose dynamical Chern numbers sum up to zero~\cite{Fu2022}. \TT{Another qualitative explanation for this zero displacement at high nonlinearity strength is that as when the nonlinearity strength $g$  increases to the point that all other terms become negligible with respect to the constant onsit potential, the pumping path becomes effectively flat and the dynamically evolved solitons appear almost constant.} Note that both methods once again yield identical results at all time, showing that the evolution of the soliton is also adiabatic in the strong nonlinearity regime.

\TT{It is important to note that even if the displacements of the soliton in the weak and strong nonlinearity regimes are quantized, the nonlinear pumping of solitons is a completely different phenomenon than linear Thouless pumping. While linear Thouless pumping assumes uniform occupation of the pumped band, this is definitely not the case when considering the pumping of a soliton, for which the localization of the wavepacket can oscillate between different sites~\cite{Fu2022}.} 

\begin{figure}
  \centering
  \includegraphics[width=0.8\columnwidth]{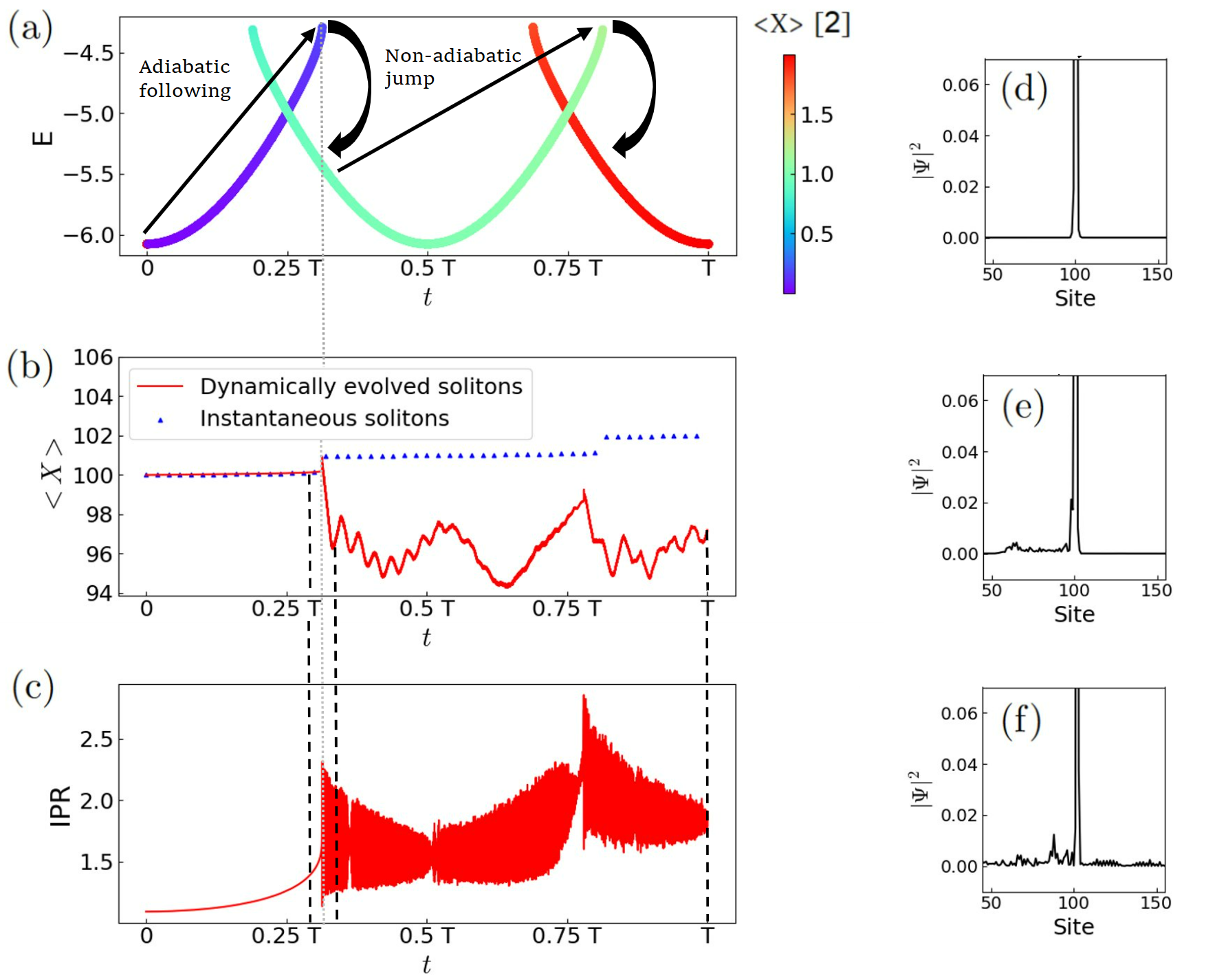}
  \caption{Study of the dynamical evolution of an initial soliton in the intermediate nonlinearity regime. Sub-figure a) shows the nonlinear energy spectrum over one cycle. Sub-figure b) compares the expectation value of position for the dynamical and instantaneous solitons over one cycle. Sub-figure c) shows the IPR of the dynamical soliton over the same cycle. Sub-figures d,e,f) show the profile of the dynamically evolved solitons at different times in the cycle. d) is at $t = 0.3 T$ right before the first jump in the trajectory of instantaneous solitons. e) is at $t = 0.34T$ right after the first jump. f) is at $t = T$ at the end of the pumping cycle. Parameters are $g = 5$ and $\omega = 10^{-3}$.}
  \label{fig:IntermediateCase}
\end{figure}

As a main observation of this work, there exists an intermediate regime where pumping is still occurring, but no longer quantized, as shown in figure~\ref{fig:IntermediateCase}-b). Over one period, the trajectory of the COM of instantaneous solitons itself is discontinuous at two points. It is precisely at these points that the dynamically evolved soliton starts to deviate from the instantaneous soliton, and its position expectation value begins to oscillate irregularly. These discontinuities in the trajectory of the COM of instantaneous solitons are tied to the observation that, as nonlinearity increases, these trajectories tend to remain close to integer positions $\langle X \rangle = n$ for longer durations, swiftly moving from one near-integer position to the next one over a short period of time, as shown later in figure~\ref{fig:TransitionWeakIntermediate}-a). This tendency culminates in the opening of gaps in the trajectory around the half-integer position $\langle X \rangle = n+\frac{1}{2}$ as instantaneous stable solitons with such positions cease to exist.

In figures~\ref{fig:IntermediateCase}-c,d,e,f), we inspect the soliton's profile and its Inverse Participation Ratio (IPR) at different time steps in the intermediate nonlinearity regime. The IPR of a state $\ket{\Psi}$ is defined by 
\begin{equation}
\label{eqn:IPR}
    \operatorname{IPR}(\ket{\Psi}) = \frac{1}{\sum\limits_{n=1}^{2N} |\Psi_n|^4},
\end{equation}
and is small (large) for localized (delocalized) states. Despite the irregular variations in the position expectation value, the dynamically evolved state actually remains strongly localized throughout the pumping cycle and ends with a strong peak at site $n=102$ after one period. Figure~\ref{fig:IntermediateCase}-c) indeed shows that the IPR remains close to the minimum value of 1 during the entire cycle. 

The seemingly random variations in the position of the COM are in fact due to the apparition of noisy perturbation as the wavepacket evolves from a soliton, instantaneous stationary state of the nonlinear Hamiltonian, to a noisier profile (imperfect soliton) as it slightly disperses to the left side of the lattice. This spreading first occurs at the time of the first jump as shown in figure~\ref{fig:IntermediateCase}-e). This perturbation ends up spreading all over the lattice as shown in figure~\ref{fig:IntermediateCase}-f) and leads to the nonquantized pumping despite the wavepacket still exhibiting a strong peak at site $n=102$ as in the weak nonlinearity regime. Therefore, despite the nonquantized value in the position of the COM, a quantized result is still obtained when solely viewing the dynamics of the wavepacket's peak. However, given that the position of the COM is directly linked to the Chern number in the linear limit, it remains a more important quantity that provides insights into the topology of the system. That this quantity is no longer quantized thus suggests a hidden mechanism that leads to a nonlinearity induced topological phase transition in the intermediate nonlinearity regime. We will uncover such a mechanism in the next section.

\begin{figure}
  \centering
  \includegraphics[width=0.8\columnwidth]{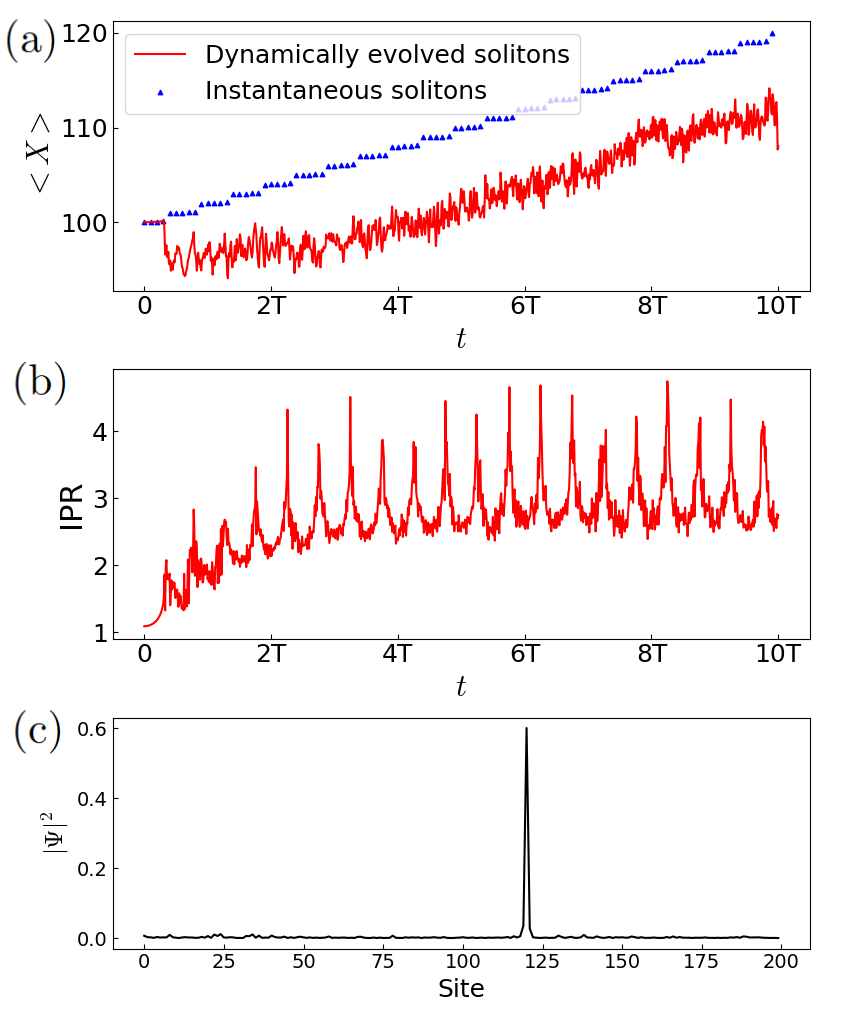}
  \caption{Dynamical evolution of an initial soliton in the intermediate nonlinearity regime over 10 pumping cycles. Sub-figure a) compares the expectation value of position for the dynamical and instantaneous solitons. Sub-figure b) shows the IPR of the dynamical soliton during the same period. Sub-figure c) shows the profile of the dynamically evolved soliton at the end of the pumping cycles at $t=10T$. Parameters are $g = 5$ and $\omega = 10^{-3}$.}
  \label{fig:LongTimeEvolution}
\end{figure}

Before ending this section, it is worth investigating the fate of the soliton over more than one pumping cycle. For many pumping cycles, the position of the COM of the dynamically evolved state follows a rightward trend, although it fails to follow the instantaneous solitons due to the perturbations generated to the left at each discontinuous jump, as shown in figure~\ref{fig:LongTimeEvolution}-a). Figure~\ref{fig:LongTimeEvolution}-b) shows that the IPR remains low with regular peaks at each discontinuity where the spreading occurs. One may notice that in between jumps the IPR decreases, meaning the wavepacket tends to relocalize around its peak. This is due to the instantaneous solitons being dynamically stable. Hence, when moved slightly away from an instantaneous soliton, the wavepacket will tend to go back to a dynamically stable instantaneous soliton. Such a small perturbation away from a stable instantaneous stationary states happens at each jump, and tends to correct itself due to the dynamical stability, until another perturbation occurs at the next jump. This is why we can finally see that even after several pumping cycles, the final state shown in figure~\ref{fig:LongTimeEvolution}-c) still has a strong peak on the expected site, i.e., $N=120$ when corresponding to 10 pumping cycles.

\section{Breakdown of the adiabatic path due to the emergence of self-crossing bands}
\label{section:Breakdown of the adiabatic due to the emergence of loop structures}

\begin{figure}
\centering
  \includegraphics[width=0.8\columnwidth]{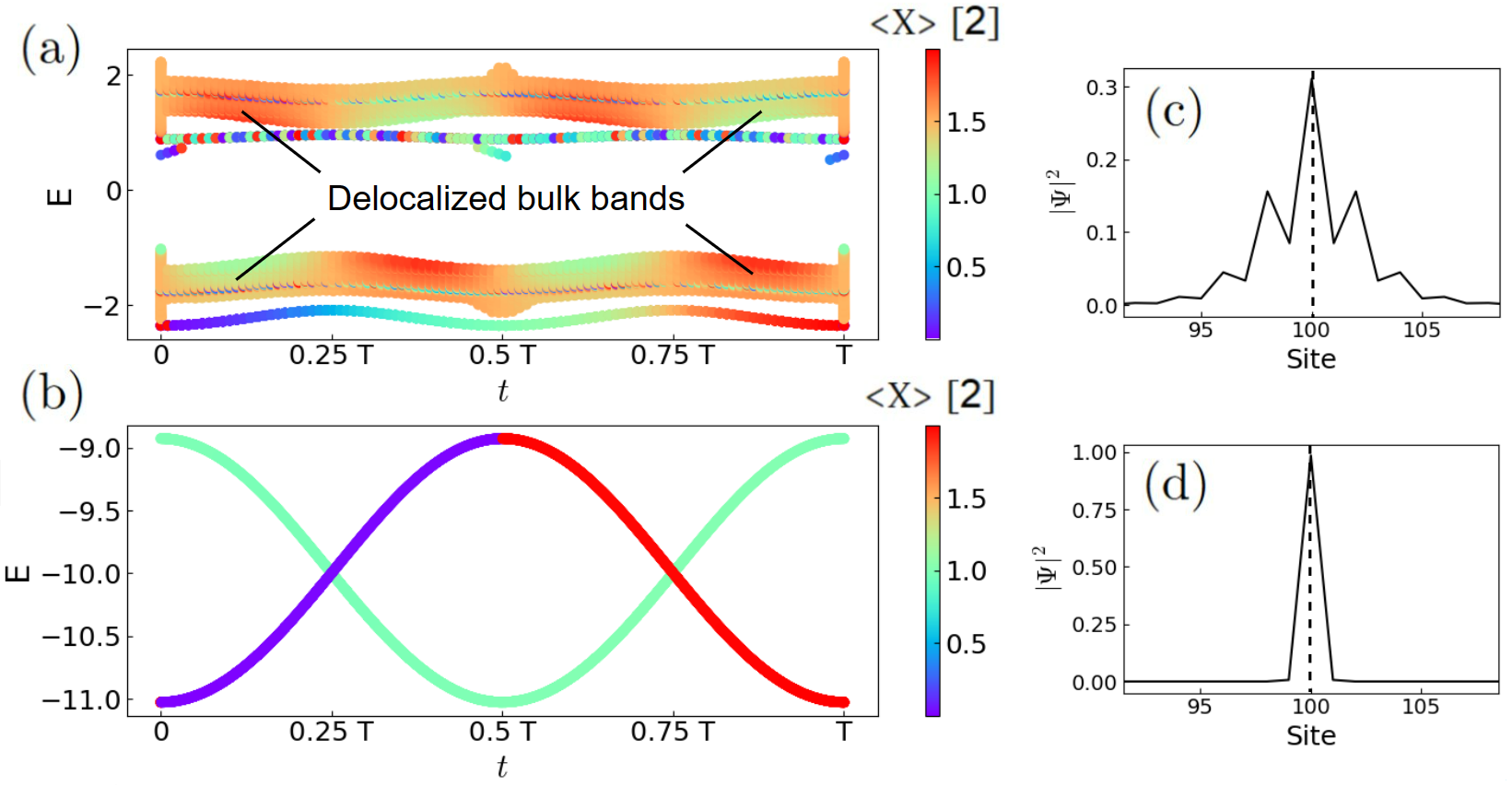}
  \caption{Nonlinear energy spectrum over one cycle for different values of nonlinearity strength, i.e., a) $g=1$ and b) $g=10$. Sub-figures c) and d) show the profile and symmetry axis of two solitons peaked at site $n=100$ at $t=0$, with nonlinearity strength of $g=1$ in c) and $g=10$ in d).}
  \label{fig:EnergySpectrumWeakStrong}
\end{figure}

The breakdown of adiabaticity in the intermediate nonlinearity regime must be investigated. To that end we examine the nonlinear energy spectrum of the system, using an iterative approach described in \ref{app:A}. In doing so we take as trial states at any given time $t$ the eigenstates of the corresponding linear model, along with the nonlinear eigenstates found at the previous time step, in order to obtain a large number of instantaneous stationary states of the nonlinear Hamiltonian and compute their energy, the results of which are shown in figure~\ref{fig:EnergySpectrumWeakStrong} for the weak and strong nonlinear regime, and in figure~\ref{fig:IntermediateCase}-a) for the intermediate nonlinear regime. The curves are color-coded by the expectation value of position of the state on the lattice modulo 2, i.e., the size of a unit cell. In the case of the strongly localized solitons that make up the majority of the energy spectrum, the symmetry of the wavefunction around its peak (as shown in figures~\ref{fig:EnergySpectrumWeakStrong}-c,d)) implies that the expectation values of position correspond to the positions of the peak of the solitons, giving their exact localization in a unit cell. Note that the energy bands corresponding to soliton states are highly degenerate, since solitons can live at any unit cell in the bulk.

In the weakly nonlinear regime of figure~\ref{fig:EnergySpectrumWeakStrong}-a), the lowest band corresponds to the pumped solitons, while the third band is another soliton band that is not involved in the pumping process. All the other bands are delocalized bulk states (with $\operatorname{IPR}(\ket{\Psi}) > 100$). The quantized pumping by one unit cell is clear from the color-coding of the band and is made possible by the continuity in the energy band that can be followed adiabatically.

\begin{figure}
\centering
  \includegraphics[width=0.8\columnwidth]{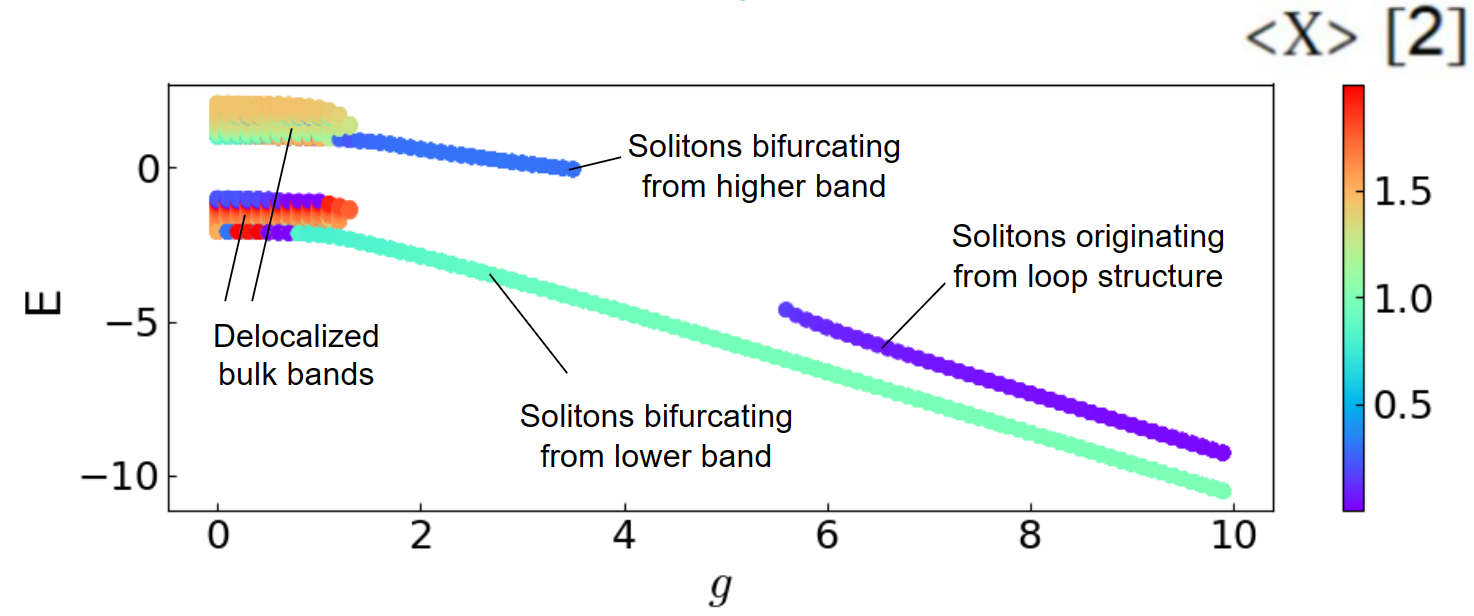}
  \caption{Nonlinear energy spectrum as a function of nonlinearity strength $g$, taken at time $t = 0.35 T$.}
  \label{fig:BifurcatingSolitons}
\end{figure}

\begin{figure}
\centering
  \includegraphics[width=\columnwidth]{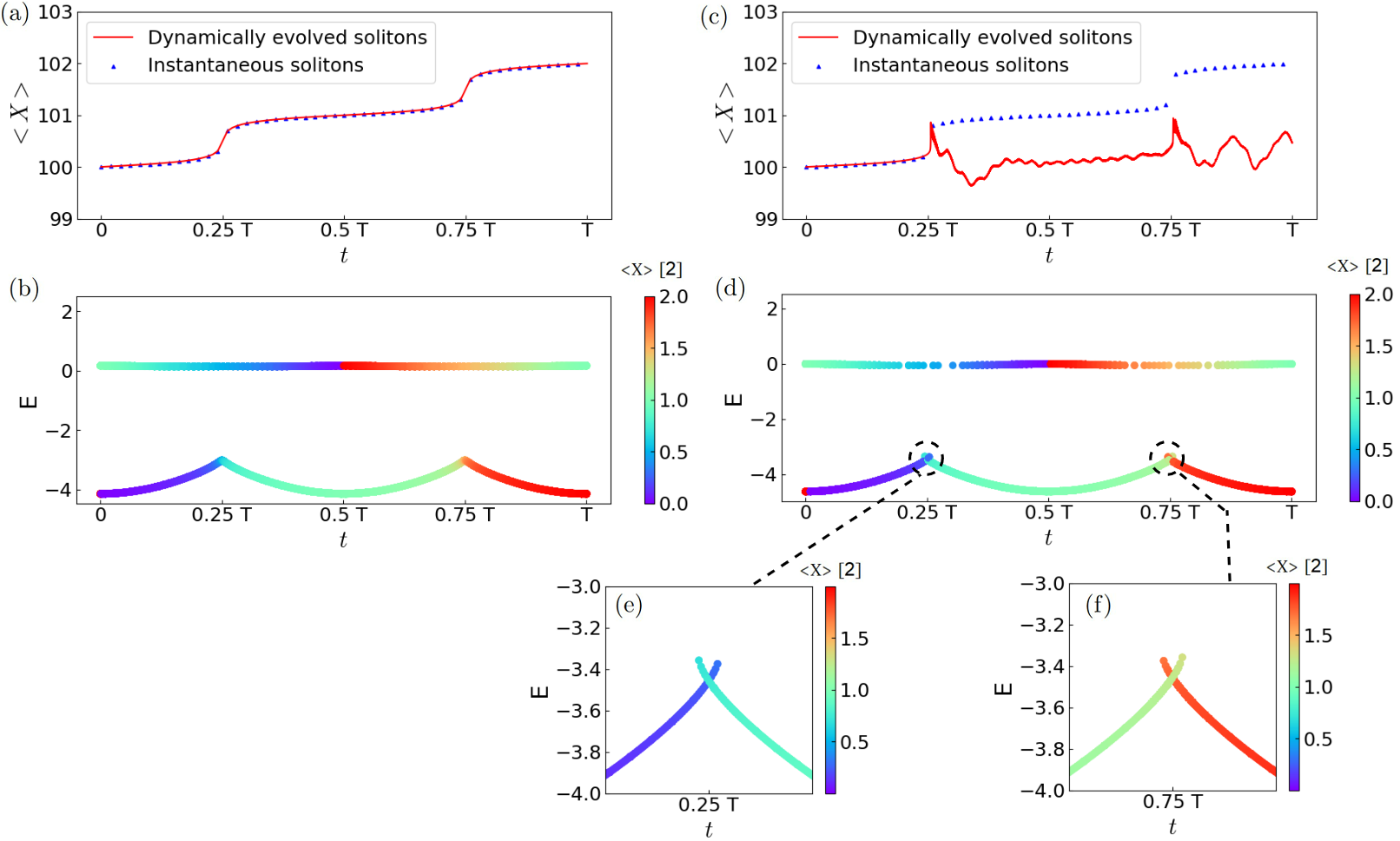}
  \caption{Nonlinear energy spectrum and dynamical pumping of solitons over one cycle for nonlinearity strengths close to the critical value $g_{crit} \approx 3.2$. Sub-figures a) and c) show the position of the instantaneous and dynamical solitons, respectively for $g=3$ and $g=3.5$. Sub-figures b) and d) show the energy spectrum for $g=3$ and $g=3.5$ respectively. For both sub-figures, the lowest band correspond to the instantaneous solitons that are pumped in a) and c). The higher, incomplete bands are also strongly localized solitons but are not considered in this pumping process. e) and f) show the apparition of the two self-crossing structures respectively around $t = 0.25 T$ and $t = 0.75 T$ for the case $g=3.5$. For dynamical evolution, the modulation frequency was taken as $\omega = 10^{-3}$.}
  \label{fig:TransitionWeakIntermediate}
\end{figure}

However, as nonlinearity strength increases, the energy spectrum of the system will undergo radical transformations, as shown in figure~\ref{fig:BifurcatingSolitons}.
The first well documented effect of the focusing nonlinear term on the system is the apparition of soliton states bifurcating from the linear bulk bands~\cite{NLZP,Jurgensen2022,Jurgensen2023}. In the system under study, one soliton band bifurcates from each bulk band, corresponding to the two soliton bands shown in figure~\ref{fig:EnergySpectrumWeakStrong}-a), the lowest one being the pumped band. As nonlinearity strength increases, the delocalized bulk bands then start to disappear due to the action of the focusing nonlinear term, that progressively favors strongly localized soliton states over delocalized states~\cite{NLZP}. The next interesting observation that can be made is that as their energy reached the level of the original linear lower bulk band, the solitons bifurcating from the higher band disappear, only leaving the pumped solitons bifurcating from the lower band. However, for even higher nonlinear strength, another soliton band appears below the energy level of the lower band. This soliton band is different in nature from the one that disappeared earlier, as it emerges from the self-crossing band, whose apparition is explained later in figure~\ref{fig:TransitionWeakIntermediate}.

Hence, the energy spectrum has already drastically changed in the intermediate nonlinearity regime of figure~\ref{fig:IntermediateCase}-a). Delocalized bulk states have already completely disappeared due to the focusing nonlinearity and the entire energy spectrum is constituted of soliton states. More strikingly, the soliton band now develops self-crossings typically observed in the energy bands of nonlinear Bloch systems~\cite{NL_LZ_Tunneling}. In this system, the self-crossing bands result in two incomplete V-shaped bands in the vicinity of $t=0.25T$ and $t=0.75T$ in figure~\ref{fig:IntermediateCase}-a). These self-crossing structures are attached to the original soliton band that spans the entire time domain and exists for all values of the nonlinearity strength, termed the main band. These incomplete V-shaped bands, along with looped-bands and swallowtail structures described in previous literature~\cite{NL_LZ_Tunneling,Diakonov2002Loop,Machholm2003Loop,Morsch2006Loop,Eckel2014Loop} collectively form a unique feature of nonlinear systems with no linear counterpart.
 
 To our knowledge, the existence of such self-crossing bands depicting the energetics of real-space localized states, let alone their effect on pumping, has not been reported before. In the strongly nonlinear regime of figure~\ref{fig:EnergySpectrumWeakStrong}-b), the two self-crossing structures merge, which causes the energy spectrum to consist of two gapless bands, each corresponding to solitons at even- and odd-numbered sites respectively. In particular, the crossing points between the two bands is responsible for the observation of Rabi-like slight oscillations in~\cite{Fu2022} and figure~\ref{fig:AdiabaticFollowingWeakStrong}-b). Note that although the existence of band crossing is known to threaten adiabaticity, the solitons belonging to the red and blue bands are strongly localized at different sublattice and hence cannot hybridize, preserving the adiabatic following throughout the cycle in figure~\ref{fig:AdiabaticFollowingWeakStrong}-c).

We will now highlight the correlation between the emergence of self-crossing bands and the breakdown of the adiabatic following. For this purpose, we study in figure~\ref{fig:TransitionWeakIntermediate} the transition between weak and intermediate nonlinearity regime, by considering the energy spectra for values of $g$ just lower and higher than the critical value $g_{crit} \approx 3.2$ at which self-crossing first appears, along with the corresponding dynamical evolution of the pumped solitons. As the system is approaching the critical value $g_{crit}$ in figure~\ref{fig:TransitionWeakIntermediate}-a) and figure~\ref{fig:TransitionWeakIntermediate}-b), the lowest energy band starts to develop cusps around which the solitons evolve at a faster rate while remaining perfectly adiabatic. However, as soon as the nonlinearity strength exceeds the critical value as in figure~\ref{fig:TransitionWeakIntermediate}-c) and figure~\ref{fig:TransitionWeakIntermediate}-d), the cusps evolve into self-crossing structures, around which the trajectory of the instantaneous solitons becomes discontinuous and adiabaticity is broken. 

The breakdown of adiabaticity in the presence of self-crossing bands can also be intuitively understood from figure~\ref{fig:IntermediateCase}. Figure~\ref{fig:IntermediateCase}-b) shows that a soliton initially (at $t = 0$) peaked on site $n=100$ evolves by adiabatically following the path formed by the blue part of the energy band in figure~\ref{fig:IntermediateCase}-a) up to $t \gtrapprox 0.25 T$, slightly moving towards site $n=101$. Ultimately, the blue band reaches a ``dead-end", beyond which there are no dynamically stable soliton solutions to adiabatically follow into. The soliton is then forced to perform a non-adiabatic jump to the green part of the energy band, which corresponds to the dynamically stable soliton with the most overlap. This in turn leads to the jump in the trajectory of instantaneous solitons, which cannot be followed by the dynamically evolved soliton. In the latter case, most of intensity of the dynamically evolved soliton will make the jump to the next suitable stable nonlinear eigenstate, while the remaining intensity will spread linearly through the lattice as shown in figure~\ref{fig:LongTimeEvolution}. A similar dead-end is observed again at $t \gtrapprox 0.75 T$, which is also accompanied by another discontinuity in the trajectory of instantaneous solitons.

On the other hand, the transition from intermediate to strong nonlinear regime can also be explained by the closing of the self-crossing bands. As nonlinear strength increases, the self-crossing structures will expand in time domain. As long as there remain incomplete bands, the ``dead-end" phenomenon leading to non-adiabatic jumps will take place leading to the non-quantized, perturbed pumping shown in figures~\ref{fig:IntermediateCase}-a,b). However, beyond a second critical value for $g$, the incomplete bands will finally spread over the entire time domain, forming the second complete band seen in figure~\ref{fig:EnergySpectrumWeakStrong}-b), marking the beginning of the strong nonlinear regime characterized by zero pumping and a recovered adiabatic following throughout the entire pumping cycle, as shown in figure~\ref{fig:AdiabaticFollowingWeakStrong}-b).

\begin{figure}
\centering
  \includegraphics[width=0.8 \columnwidth]{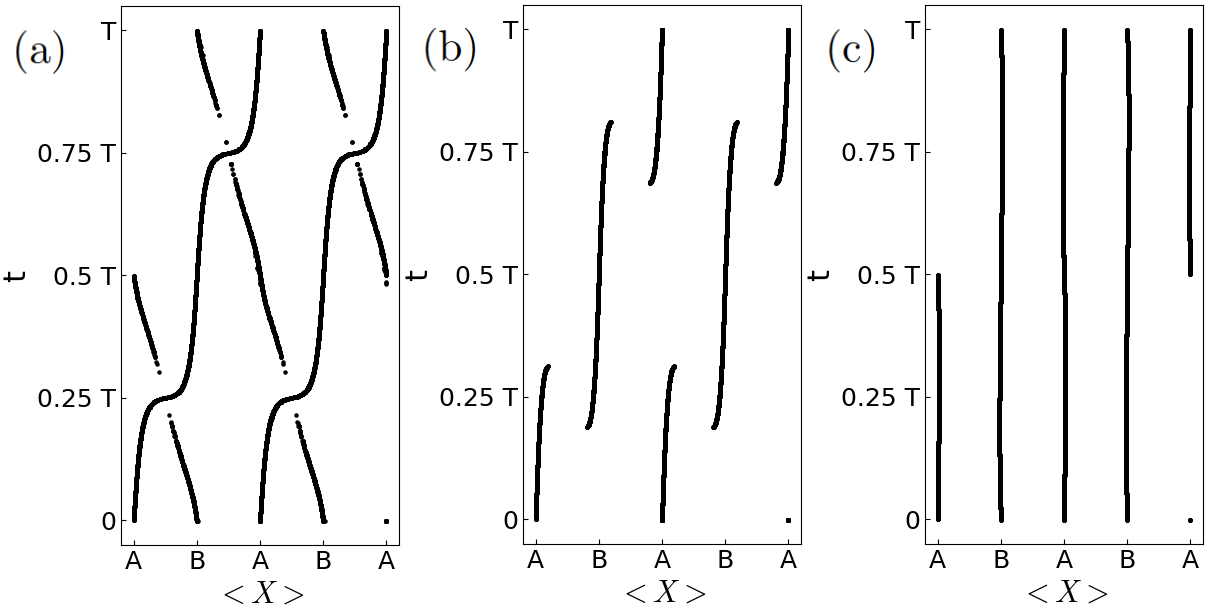}
  \caption{Position of the center of mass of all available instantaneous solitons over one pumping cycle, for nonlinearity strength of a) $g=3$, b) $g=5$ and c) $g=10$. Letters A and B correspond respectively to even and odd lattice sites.}
  \label{fig:Trajectories}
\end{figure}

The effect of self-crossing bands on soliton pumping can be clearly understood by considering all possible trajectories of instantaneous solitons over one pumping cycle as shown in figure~\ref{fig:Trajectories}. Figures~\ref{fig:Trajectories}-a,c) respectively correspond to the weak and strong nonlinear regime, where quantized pumping is possible and was experimentally observed~\cite{Jurgensen2021ExperimentalSolitonPumping} thanks to the contiguous path that can be adiabatically followed, leading respectively two sites away or to the same site after one pumping cycle. In the intermediate nonlinear regime shown in figure~\ref{fig:Trajectories}-b), the apparition of self-crossing bands coincide with the disappearance on any contiguous path making adiabatic following impossible. Although this intermediate regime only occurs over a very small window for the system considered in reference \cite{Jurgensen2021ExperimentalSolitonPumping}, it is an important transition regime showing the failure of adiabaticity due to nonlinear effects.

The analysis of the intermediate nonlinearity regime, characterized by the presence of self-crossing bands in the real space energy spectrum and the breakdown of the adiabatic following due to the dead-ends in the adiabatic path, plays the role of missing link between the two other, more extensively studied, regimes. It allows us to draw the bridge between the weakly nonlinear regime where adiabatic following is possible and pumping of the soliton remains quantized~\cite{Jurgensen2022,mostaan2022quantized,Fu2022}, and the strongly nonlinear regime, where solitons oscillate around the same state, amounting to a zero total displacement~\cite{Fu2022}. Moreover, the correspondence between the presence of self-crossing bands and the breakdown of quantized pumping established here provides a means of detecting the former via an adiabatic pumping experiment.

\section{Nonlinear adiabatic pumping of Bloch states}
\label{section:Nonlinear adiabatic pumping in momentum space}

To gain a different perspective into the breakdown of adiabaticity and nonquantized pumping due to self-crossing bands, we will now consider a different type of pumping involving Bloch states rather than a localized soliton. To this end, we consider the model under Periodic Boundary Conditions (PBC), and assume Bloch state solutions
\begin{eqnarray}
    \label{eqn:PhaseSpaceModel}
    \Psi_{2j} = \Phi_{e} e^{\rmi k j}\\
    \Psi_{2j+1} = \Phi_{o} e^{\rmi k j}.
\end{eqnarray}

 At any given time, we can solve the system for instantaneous stationary states by solving the nonlinear eigenvalue problem $H(t,k,\ket{\Phi(t,k)}) \ket{\Phi(t,k)} = E \ket{\Phi(t,k)}$, where $\ket{\Phi} = [\Phi_e,\Phi_o]^{T}$ is a pseudo-spinor and $H(t,k,\ket{\Phi(t,k)})$ is the instantaneous two-band Gross–Pitaevskii (GP) Hamiltonian
\begin{equation}
    H(k,t,\ket{\Phi(k,t)}) = h_x(k,t) \, \sigma_x + h_y(k,t) \, \sigma_y + h_z(t) \, \sigma_z  - g\begin{pmatrix} \left|\Phi_{e}(k,t)\right|^2 & 0 \\ 0 &  \left|\Phi_{o}(k,t)\right|^2 \end{pmatrix},
    \label{eqn:GPHamiltonian}
\end{equation}
 where $\sigma$'s are the Pauli matrices in the standard representation, $h_x(k,t) = - [J + \delta \sin{\omega t}] - [J - \delta \sin{\omega t}] \cos{k}$, $h_y(k,t) =  - [J - \delta \sin{\omega t}] \sin{k}$, and $h_z(t) = - \Delta \cos{\omega t}$. By defining $\Sigma = \left|\Phi_{o}\right|^2 - \left|\Phi_{e}\right|^2$ as the population difference between the two pseudo-spinor components, we can rewrite the Hamiltonian in a more compact form
 \begin{equation}
    H(k,t,\Sigma) = h_x(k,t) \, \sigma_x + h_y(k,t) \, \sigma_y + h(t,\Sigma) \, \sigma_z - \frac{g}{2} I_2,
    \label{eqn:CompactGPHamiltonian}
\end{equation}
where $h(t,\Sigma) = h_z(t) + \frac{g}{2}\Sigma$ and $I_2$ is a $2\times 2$ identity matrix that will be ignored in later calculations as it merely contributes a shift of the energy. 

\begin{figure}
\centering
  \includegraphics[width=0.8\columnwidth]{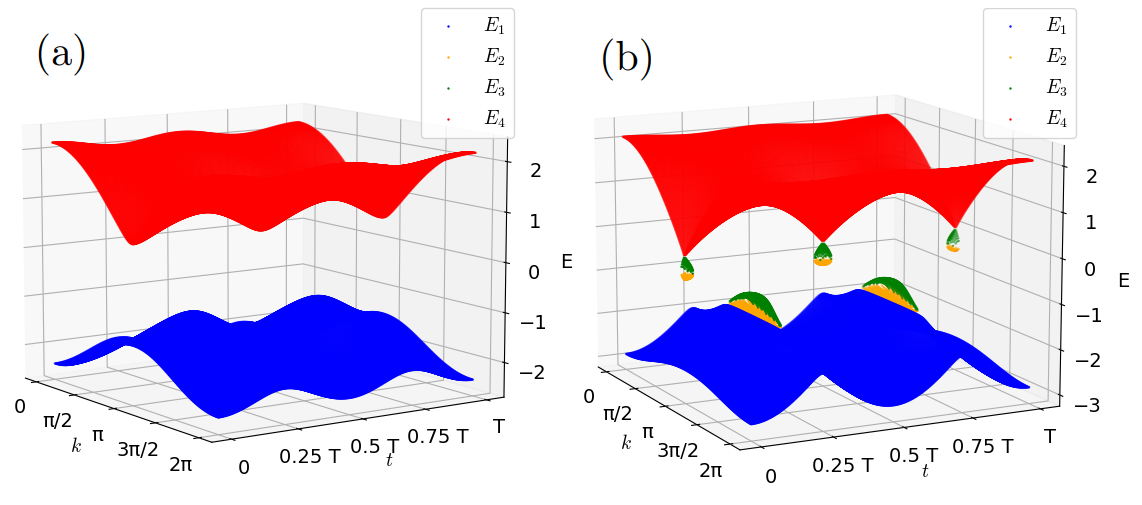}
  \caption{Nonlinear energy bands of the Bloch state pumping model, for different values of the nonlinearity strength $g$. a) is $g=1$ and b) is $g=3$.}
  \label{fig:BlochBands}
\end{figure}

In the corresponding linear model, the system is gapped at all times for all quasimomenta, allowing for adiabatic pumping of particles. As shown in figure~\ref{fig:BlochBands}-a), a similar gapped band structure is observed at weak nonlinearity. However, at intermediate nonlinearity, i.e., $g > g_c = 2$, the nonlinear energy spectrum starts to sport self-crossing bands, which this time form fully looped structures made of a pair of incomplete energy bands that do not span the whole two-dimensional Brillouin Zone, see e.g. the orange- and green-colored bands in figure~\ref{fig:BlochBands}-b). We will now analytically show that the development of loop structures plays a significant role in modifying the pumping of such a Bloch state, this time without breaking the adiabaticity of the process.

Recall that the displacement after one pumping cycle can be written as 
\begin{equation}
    \Delta \left< x \right> = \frac{1}{2 \pi} \int \, \rmd t \int_{-\pi}^{\pi} \, \rmd k \left< \frac{\partial H}{\partial k} \right> ,  
    \label{eqn:Delta}
\end{equation}
where $\left< ... \right>$ represents the average over the state at a given $(k,t)$. During adiabatic evolution, this state is known to only slightly deviate from the instantaneous stationary state by a quantity whose average yields the Berry curvature. However, we show in \ref{app:B} that the presence of nonlinear terms causes the state to deviate further from the stationary state, resulting in an additional contribution which averages in a nonlinear correction to the Berry curvature. For the system given in equation~(\ref{eqn:CompactGPHamiltonian}) and a state $\Phi = (\cos{[\theta(k,t)/2]},\sin{[\theta(k,t)/2]} \exp[i \phi(k,t)])^{T}$, the average displacement over one adiabatic cycle is given by
\begin{equation}
    \begin{aligned}
    \Delta \left< x \right> &= \frac{1}{2 \pi} \int  \rmd t \int_{-\pi}^{\pi}  \rmd k  \left[ \mathcal{B}(k,t) + \mathcal{D}(k,t) \right] \;,
    \end{aligned}
    \label{eqn:Displacement}
\end{equation}
where $\mathcal{B}(k,t)$ is the Berry curvature and
\begin{equation}
    \mathcal{D}(k,t) = - \frac{g \sin^3{\theta}}{2E + g \sin^2{\theta}} \frac{\partial \phi}{\partial t} \frac{\partial}{\partial k} \left( \frac{\theta}{2} \right) 
    \label{eqn:Drift}
\end{equation}
is the nonlinear correction to the Berry curvature causing the average displacement to drift away from the integral of the conventional Berry curvature. This shows that unlike linear Thouless pumping, the displacement of the nonlinear adiabatic pumping is not simply proportional to the Chern number due to this extra second term. \TT{It is however much closer \RM{conceptually} to the linear Thouless pumping than the soliton pumping, as we also assume an averaged displacement over all possible Bloch waves, so that it reduces to the standard linear pumping in the linear limit $g = 0$.} 
 
\begin{figure}
\centering
  \includegraphics[width=0.8\columnwidth]{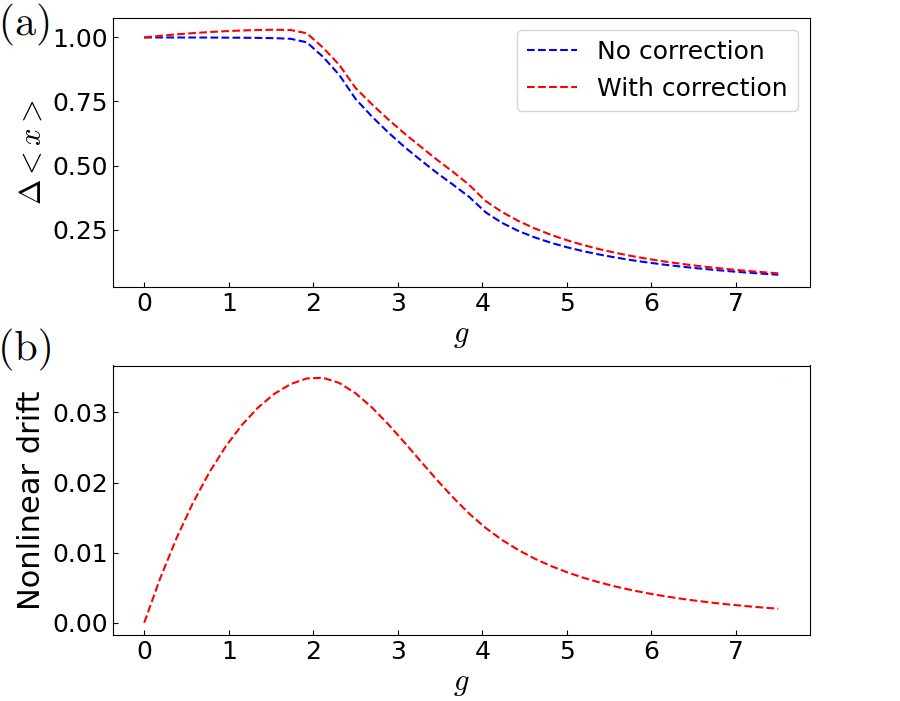}
  \caption{Panel a) shows the average displacement of a particle over one adiabatic cycle, for different values of nonlinearity strength. The red and blue lines respectively show the displacement with and without the additional nonlinear drift. Panel b) shows the nonlinear drift alone.}
  \label{fig:Displacement}
\end{figure}

We then compute the average displacement over one adiabatic cycle for the lowest band of our system, for different values of the nonlinear strength. Results are shown on figure~\ref{fig:Displacement}. Note that although the result presented above is derived for the highest band of the two band system by choosing $\ket{+} = (\cos{(\theta/2)},\sin{(\theta/2)} e^{\rmi \phi})^{T}$ as our initial state, it can easily be adapted to the lowest band with initial state $\ket{-} = (\sin{(\theta/2)},- \cos{(\theta/2)} e^{\rmi \phi})^{T}$ by simply replacing $\theta$ by $\theta' = \theta - \pi$ in all expressions. 

The variations of the displacement with the nonlinearity strength shown in figure~\ref{fig:Displacement}-a) present many interesting features. First for weak nonlinearity strength, the pumping occurs but is no longer quantized to the Chern number due to the additional nonlinear drift. Moreover, in the strongly nonlinear limit, the average displacement goes to zero, with or without the additional nonlinear drift. 
Interestingly, the transition between a quantized displacement and a zero displacement starts exactly at $g = g_c = 2$, when the loop structures first appear. Beyond this critical value, the displacement is no longer quantized even without taking the nonlinear drift into account, and starts continuously decreasing to zero. Intuitively, this is attributed to the fact that in the presence of loop structures, the nonlinear energy bands are no longer gapped and their Chern number is consequently no longer quantized.

Considering the additional nonlinear drift alone also offers some interesting observations as shown in figure~\ref{fig:Displacement}-b). Starting small at weak nonlinearity, it reaches a peak value, then becomes small again at strong nonlinearity. Specifically, the peak value occurs exactly at the critical value $g_c$ where the loop structures start to appear. Coincidentally, the critical value $g_c$ corresponds to the first apparition of singular points in the expression of both the Berry curvature, and its nonlinear correction. These results are shown in detail in \ref{app:C}.

\section{Discussion}
\label{section:Discrepancy between real space and momentum space pumping}

While the analysis presented in section~\ref{section:Nonlinear adiabatic pumping in momentum space} highlights the direct correlation between self-crossing bands and nonquantized pumping, it is to be stressed that the results therein may not have a direct correlation with our numerical results presented in section~\ref{section:Nonlinear Thouless pump} and \ref{section:Breakdown of the adiabatic due to the emergence of loop structures}. Indeed, as elucidated before, the former represents pumping of Bloch states, whereas the latter describes pumping of localized solitons.

The two pumping scenarios above only reconcile in the absence of nonlinearity due to the momentum integral in equation~(\ref{eqn:Delta}), which effectively turns the various Bloch states into a localized Wannier state. Since superposition principle no longer holds in the nonlinear setting, equation~(\ref{eqn:Delta}) is in general inequivalent to the displacement of the particle over one pumping cycle at finite nonlinearity. This in turn explains the order of magnitude difference between the nonlinear correction analytically derived in section~\ref{section:Nonlinear adiabatic pumping in momentum space} and that numerically observed in section~\ref{section:Nonlinear Thouless pump} and \ref{section:Breakdown of the adiabatic due to the emergence of loop structures}. Nevertheless, it is remarkable to highlight that the pattern of increasing discrepancy between dynamically evolved solitons and instantaneous solitons as the self-crossign structures grow holds in both types of pumping. \TT{This demonstrates that our finding that self-crossing bands, which are a purely nonlinear phenomenon, can lead to the breakdown of quantized pumping in spite of the topological stability of the system holds for a general type of pumping.}

It is also worth noting the difference in normalization prescription for the two pumping treatments. That is, in the soliton pumping of sections~\ref{section:Nonlinear Thouless pump} and \ref{section:Breakdown of the adiabatic due to the emergence of loop structures}, the wave function is normalized over the entire lattice, i.e., $\sum_{n=1}^{2N} |\Psi_n|^2 = 1$. By contrast, in the Bloch states pumping of section~\ref{section:Nonlinear adiabatic pumping in momentum space}, the wave function is only normalized over a unit cell, i.e., $|\Phi_e|^2  + |\Phi_o|^2 =1$. This leads to different scaling of the nonlinearity strength $g$ which, among others, contributes to the discrepancy between the critical nonlinearity values in the two pumping treatments. Establishing a quantitative comparison between the two pumping treatments thus remains a challenging question and will be left for future work.

\section{Concluding remarks}
\label{section:Concluding remarks}

In this work, we have studied the correlation between the pumping of solitons and the nonlinear energy bands of a system for a wide range of nonlinearity strength. We established the missing link between earlier studies of~\cite{Jurgensen2022,mostaan2022quantized,Fu2022,Jurgensen2023} by identifying an intermediate nonlinearity regime in which the pumping is nonzero but no longer quantized. This phenomenon is attributed to the presence of self-crossing bands in the nonlinear energy band, disrupting the trajectory of the evolved solitons. Finally, we analytically quantify the nonlinear correction to displacement in the case of Bloch state pumping, which further quantifies the effect of such self-crossing bands. In another recent work, we explicitly utilize this nonlinear correction to probe nonlinear effects on Weyl semimetals~\cite{NonlinearWeylSemimetal}.

In a future study, it would be interesting to modify the adiabatic driving, e.g., in the spirit of the driven nonlinear Landau-Zener process presented in~\cite{Liu2002LandeauZener,Zhang2008}, to bypass the self-crossings. This would enable a successful (quantized) adiabatic pumping for a larger range of nonlinearity, extending it to the intermediate nonlinearity regime, and possibly even in the strongly nonlinear regime.

\ack
Part of the work done by R.W.B was supported by the Australian Research Council Centre of Excellence for Engineered Quantum Systems (EQUS, CE170100009). R.W.B additionally acknowledges the support provided by the Deanship of Research Oversight and Coordination (DROC) at King Fahd University of Petroleum \& Minerals (KFUPM) through project No.~EC221010. J.G. is funded by the Singapore National Research Foundation Grant No. NRF-NRFI2017-04 (WBS No. R-144-000-378- 281).

\appendix

\section{Self-consistent iterative method for instantaneous stationary states} 
\label{app:A}

In order to find instantaneous stationary states of the nonlinear lattice at any given time, we use an iterative method. For a given nonlinear, state dependent Hamiltonian $H$, at a given time $t$, the iterative process from a state $\ket{\Psi_n(t)}$ to the state $\ket{\Psi_{n+1}(t)}$ is as follows:
\begin{itemize}
    \item We first compute $H_n = H(\ket{\Psi_n},t)$, the nonlinear state-dependent Hamiltonian evaluated at the state $\ket{\Psi_n(t)}$.
    \item We then solve $H_n$ for its eigenstates $\ket{\psi_i}$ with $i = 1,...,2N$.
    \item We finally choose the new state $\ket{\Psi_{n+1}(t)}$ as the one eigenstate $\ket{\psi_i}$ with the largest overlap with the previous $\ket{\Psi_n(t)}$. More specifically, we take $\ket{\Psi_{n+1}(t)} = \ket{\psi_{i_0}}$ where $| \langle \Psi_n(t) | \psi_{i_0}\rangle | \geq | \langle \Psi_n(t) | \psi_{i}\rangle | $ for all $i$.
\end{itemize}
we repeat this iterative process until the overlap between old and new state is close enough to 1 by an arbitrary $\epsilon$, i.e. $ | \langle \Psi_{n+1}(t) | \Psi_{n}(t) \rangle | > 1 - \epsilon$. Throughout this work, we take $\epsilon = 10^{-10}$. 
In order to execute this iteration method, one also needs to choose the initial state used as a starting point of the iteration. Note that the stationary state obtained is quite dependent on the chosen starting point (trial state). In order to obtain instantaneous solitons, we found that using sech states 
\begin{equation}
\label{eqn:A_SechState}
\begin{aligned}
\Psi_n &\propto \cosh{\left(\frac{|n-100|}{5}\right)}^{-1} \\
\sum_{n=1}^{2N} &|\Psi_n|^2 = 1.
\end{aligned}
\end{equation}
as trial states gives the best results~\cite{Narita1990SechSoliton,Narita1991SechSoliton,Maruno2006SechSoliton,Wang2014SechSoliton}. To obtain the nonlinear energy spectra, we also used eigenstates of the corresponding linear model as trial states, allowing us to obtain delocalized bulk states in the weakly nonlinear regime.

\section{Displacement correction in Bloch state pumping}
\label{app:B}

We consider a general two level GP Hamiltonian
\begin{equation}
\label{eqn:B_GPHamiltonian}
    H(\ket{\Phi}) = h_x \sigma_x + h_y \sigma_y + h(\Sigma) \sigma_z 
\end{equation}
where $\Sigma = \abs{\Phi_2}^2 - \abs{\Phi_1}^2$. We assume that $H,h_x,h_y,h$ are all functions of a quasimomentum $k$ and a periodic time modulation $\omega t$ where $\omega$ is the frequency of said modulation, always appearing with $t$. We define a state $\Psi_a = e^{-if} \Phi_a$ with $a = 1,2$, which corresponds to an element of a projective Hilbert space \cite{Liu2010NLBerryPhase}. The total phase $f$ is taken to capture both dynamical and geometric phases of the state $\ket{\Phi}$. With this notation, the nonlinear Schrödinger equation now reads (summation of repeated indices being implied)
\begin{equation}
    \label{eqn:B_dfdt}
    \frac{\rmd f}{\rmd t} \Psi_a = \rmi \frac{\rmd \Psi_a}{\rmd t} - H_{ab} \Psi_b.
\end{equation}
Applying $\sum_{a} \Psi_a^{*} ...$ to both sides, we obtain 
\begin{equation}
    \frac{\rmd f}{\rmd t} = \rmi \Psi_a^{*} \frac{\rmd \Psi_a}{\rmd t} - \Psi_a^{*} H_{ab} \Psi_b .
\end{equation}
Doing a perturbative expansion of $f$ and $\Psi_a$ under an adiabatic parameter $\epsilon$ gives
\begin{equation}
    \label{eqn:B_expansion}
    \begin{aligned}
    \frac{\rmd f}{\rmd t} &= \alpha_0 + \alpha_1 \epsilon + ... \\
    \Psi_a &= \Psi_a^{(0)} + \epsilon \Psi_a^{(1)} + ...    
    \end{aligned}
\end{equation}
and since the nonlinear Hamiltonian is also state dependent, we will also have
\begin{equation}
\label{eqn:PerturbativeExpansionHamiltonian}
    H = H^{(0)} + \epsilon H^{(1)} +  ...
\end{equation}

We now attempt to derive the small correction $\epsilon \Psi_a^{(1)}$ away from the stationary state $\Psi^{(0)} = \ket{+}$ such that $H^{(0)} \ket{+} = E \ket{+}$ during the adiabatic process. Separating between zeroth and first order terms in equation~(\ref{eqn:B_expansion}), we obtain
\begin{equation}
\label{eqn:B_GeneralAdiabaticLimit}
    \begin{aligned}
    \alpha_0 &= -E,\\
    \epsilon \alpha_1 &= \rmi \Psi_a^{(0)*} \frac{\rmd \Psi_a^{(0)}}{\rmd t} - \epsilon\Psi_a^{(0)*} H_{ab}^{(1)} \Psi_a^{(0)},
    \end{aligned}
\end{equation}
where we notice that the first term in the right hand side of the bottom line corresponds to the conventional Berry connection, and the second term is the geometric contribution coming from the dynamical phase, due to nonlinear dynamics. In our case, we have $H^{(1)} = \left. \frac{\rmd h}{\rmd \Sigma} \right\rvert_{\Sigma = \Sigma^{(0)}} \left. \frac{\rmd \Sigma}{\rmd \epsilon} \right\rvert_{\epsilon = 0} \sigma_z $. Using the normalization condition $\operatorname{Re}(\Psi_a^{(0)*} \Psi_a^{(1)}) = 0$, we have $\left. \frac{\rmd \Sigma}{\rmd \epsilon} \right\rvert_{\epsilon = 0} =  - 4 \operatorname{Re}(\Psi_1^{(0)*} \Psi_1^{(1)})$ so
\begin{equation}
    \label{eqn:H1Perturbation}
    \begin{aligned}
    H^{(1)}= -4 \left. \frac{\rmd h}{\rmd \Sigma} \right\rvert_{\Sigma = \Sigma^{(0)}} \operatorname{Re}(\Psi_1^{(0)*} \Psi_1^{(1)}) \sigma_z . 
    \end{aligned}
\end{equation}
The general formula for $\alpha_0$ and $\alpha_1$ given in equation(\ref{eqn:B_GeneralAdiabaticLimit}) becomes then
\begin{equation}
		\label{eqn:B_Alpha0and1}
            \begin{aligned}
            \alpha_0 &= -E  \\
             \epsilon \alpha_1 &= \rmi \Psi_a^{(0)*} \frac{\rmd \Psi_a^{(0)}}{\rmd t} - 4 \epsilon \left. \frac{\rmd h}{\rmd \Sigma} \right\rvert_{\Sigma^{(0)}} \Sigma^{(0)} \operatorname{Re}(\Psi_1^{(0)*} \Psi_1^{(1)}) .
            \end{aligned}
\end{equation}
On the other hand, if we consider only $\epsilon^1$ terms in equation~(\ref{eqn:B_dfdt}), using equation~(\ref{eqn:B_Alpha0and1}) we have for $a=1$
\begin{multline}
\label{eqn:B_InjectInGPa=1}
    4 \epsilon \left. \frac{\rmd h}{\rmd \Sigma} \right\rvert_{\Sigma^{(0)}} \operatorname{Re}(\Psi_1^{(0)*} \Psi_1^{(1)}) [\Sigma^{(0)}+1] \Psi_1^{(0)} = \\ -\rmi (\delta_{1a} - \Psi_1^{(0)} \Psi_a^{(0)*})\frac{\rmd \Psi_a^{(0)}}{\rmd t} - \epsilon (E \delta_{1b} - H_{1b}^{(0)}) \Psi_b^{(1)} .
\end{multline}
and for $a=2$
\begin{multline}
\label{eqn:B_InjectInGPa=2}
    4 \epsilon \left. \frac{\rmd h}{\rmd \Sigma} \right\rvert_{\Sigma^{(0)}} \operatorname{Re}(\Psi_1^{(0)*} \Psi_1^{(1)}) [\Sigma^{(0)}-1] \Psi_2^{(0)} = \\ -\rmi (\delta_{2a} - \Psi_2^{(0)} \Psi_a^{(0)*})\frac{\rmd \Psi_a^{(0)}}{\rmd t} - \epsilon (E \delta_{2b} - H_{2b}^{(0)}) \Psi_b^{(1)} .
\end{multline}

For a two-level system, the stationary state $\ket{+}$ can be written without loss of generality in the form
\begin{equation}
\label{eqn:B_StationaryPsiE}
    \ket{+}=\begin{pmatrix} \cos{\frac{\theta}{2}} \\ \sin{\frac{\theta}{2}}e^{\rmi\phi} \end{pmatrix},
\end{equation}
so that we can simplify equation~(\ref{eqn:B_InjectInGPa=1}) by taking its real part, and making use again of the normalization condition $\cos{\frac{\theta}{2}} \operatorname{Re}(\Psi_1^{(1)}) + \sin{\frac{\theta}{2}} \operatorname{Re}(e^{-\rmi \phi} \Psi_2^{(1)}) = 0$ to get
\begin{multline}
\label{eqn:B_RealPartInjectInGP}
    4 \epsilon \left. \frac{\rmd h}{\rmd \Sigma} \right\rvert_{\Sigma^{(0)}} \cos^2{\frac{\theta}{2}} [1 - \cos{\theta}] \operatorname{Re}( \Psi_1^{(1)})  = \rmi \cos{\frac{\theta}{2}} \Psi_a^{(0)*}\frac{\rmd \Psi_a^{(0)}}{\rmd t} \\ - \epsilon (E - H_{11}^{(0)} + \cot{\frac{\theta}{2}} H_{12}^{(0)} e^{\rmi \phi}) \operatorname{Re}(\Psi_1^{(1)}) .
\end{multline}
Now we can notice that 
\begin{equation}
		\label{eqn:B_HiddenEigenState}
            \begin{aligned}
            \ket{+^{\perp}} = \begin{pmatrix} \sin{\frac{\theta}{2}} \\ -\cos{\frac{\theta}{2}} e^{\rmi\phi} \end{pmatrix}
            \end{aligned}
\end{equation}
is the (hidden) eigenstate \footnote{$\ket{+^{\perp}}$ is however not a stationary state of the system, as $H^{(0)}$ is state dependent, and $\ket{+^{\perp}}$ is an eigenstate of $H^{(0)}\left(\ket{+}\right)$, but not necessarily of $H^{(0)}\left(\ket{+^{\perp}}\right)$} of $H^{(0)}$ with eigenvalue $ - E$ to simplify the last line in equation~(\ref{eqn:B_RealPartInjectInGP}). We then obtain
\begin{equation}
    \label{eqn:B_RealPartofPsi1}
    \epsilon \operatorname{Re}(\Psi_1^{(1)}) = \frac{\cos{\frac{\theta}{2}}}{2E + 2 \left. \frac{\rmd h}{\rmd \Sigma} \right\rvert_{\Sigma^{(0)}} \sin^2{\theta}} \rmi \Psi_a^{(0)*}\frac{\rmd \Psi_a^{(0)}}{\rmd t}.
\end{equation}
Using the normalization condition, we also get $\operatorname{Re}(e^{-\rmi \phi} \Psi_2^{(1)})$ as
\begin{equation}
    \label{eqn:B_RealPartofPsi2}
    \epsilon \operatorname{Re}(e^{-\rmi \phi} \Psi_2^{(1)}) = \frac{-  \cot{\frac{\theta}{2}}  \cos{\frac{\theta}{2}}}{2E + 2 \left. \frac{\rmd h}{\rmd \Sigma} \right\rvert_{\Sigma^{(0)}} \sin^2{\theta}} \rmi \Psi_a^{(0)*}\frac{\rmd \Psi_a^{(0)}}{\rmd t}.
\end{equation}

In order to determine their imaginary part, we now take the imaginary part of equation~(\ref{eqn:B_InjectInGPa=1}) and equation~(\ref{eqn:B_InjectInGPa=2}) to get, respectively,
\begin{equation}
    \label{eqn:B_Imaginarya=1}
    -\frac{\rmd \Psi_1^{(0)}}{\rmd t} = \epsilon (E - H_{11}^{(0)}) \operatorname{Im}(\Psi_1^{(1)}) - \epsilon H_{12}^{(0)} e^{\rmi \phi}  \operatorname{Im}(e^{-\rmi \phi} \Psi_2^{(1)}),
\end{equation}
and
\begin{multline}
    \label{eqn:B_Imaginarya=2}
    - \operatorname{Im}(\rmi e^{-\rmi \phi} \frac{\rmd \Psi_2^{(0)}}{\rmd t}) = \epsilon (E - H_{22}^{(0)}) \operatorname{Im}(e^{-\rmi \phi} \Psi_2^{(1)}) - \epsilon H_{21}^{(0)} e^{-\rmi \phi}  \operatorname{Im}(\Psi_1^{(1)}),
\end{multline}
which can be more compactly written as
\begin{equation}
    \label{eqn:B_ImaginaryMatrix}
    - \operatorname{Im}\left(\rmi U \begin{bmatrix} \frac{\rmd}{\rmd t}\Psi_1^{(0)} \\ \frac{\rmd }{\rmd t}\Psi_2^{(0)} \end{bmatrix} \right) = \epsilon \operatorname{Im} \left( U (E - H^{(0)}) \begin{bmatrix} \Psi_1^{(1)} \\ \Psi_2^{(1)} \end{bmatrix}  \right)
\end{equation}
where
\begin{equation}
    U = \begin{pmatrix} 1 & 0 \\ 0 & e^{-\rmi \phi} \end{pmatrix}.
\end{equation}
Now using the eigenstate $\ket{\Psi^{(0)}} = \ket{+}$ with energy $E$ and the hidden eigenstate $\ket{\Psi^{(0) \perp}} = \ket{+^{\perp}}$ with energy $E_{\perp} = -E$, we can rewrite $H^{(0)}$ as
\begin{equation}
    \label{eqn:B_SpectralDecomposition}
    H^{(0)} = E \ket{\Psi^{(0)}} \bra{\Psi^{(0)}} - E \ket{\Psi^{(0) \perp}} \bra{\Psi^{(0) \perp}},
\end{equation}
and using the fact that $\ket{\Psi^{(0)}} \bra{\Psi^{(0)}} + \ket{\Psi^{(0) \perp}} \bra{\Psi^{(0) \perp}} = 1$ we can rewrite equation~(\ref{eqn:B_ImaginaryMatrix}) as
\begin{equation}
    \label{eqn:B_ImaginaryMatrix2}
    - \operatorname{Im}\left(\rmi U \frac{\rmd}{\rmd t}\ket{\Psi^{(0)}} \right) =  \epsilon \operatorname{Im} \left( 2 E U \ket{\Psi^{(0) \perp}} \bra{\Psi^{(0) \perp}} \ket{\Psi^{(1)}}  \right)
\end{equation}
Since $\ket{\Psi^{(1)}}$ is defined as a small correction to $\ket{\Psi^{(0)}}$ we have a freedom to make it orthogonal to $\ket{\Psi^{(0)}}$ (components of $\ket{\Psi^{(1)}}$ parallel to $\ket{\Psi^{(0)}}$ can be separated and combined with $\ket{\Psi^{(0)}}$, which only affects the global phase of $\ket{\Psi^{(0)}}$). In such a case, the projection of $\ket{\Psi^{(1)}}$ onto $\ket{\Psi^{(0) \perp}}$ will not change anything, and we finally find
\begin{equation}
\begin{aligned}
    - \operatorname{Im}\left(\rmi U \frac{\rmd}{\rmd t}\ket{\Psi^{(0)}} \right) &=  \epsilon \operatorname{Im} \left( 2 E U \ket{\Psi^{(1)}}  \right) \\
    \epsilon \operatorname{Im} \left( U \ket{\Psi^{(1)}} \right) &= \frac{-1}{2E} \operatorname{Im}\left(\rmi U \frac{\rmd}{\rmd t}\ket{\Psi^{(0)}} \right) \\
    \epsilon \operatorname{Im} \left( U \ket{\Psi^{(1)}} \right) &=  - \operatorname{Im}\left(\rmi U \frac{\bra{\Psi^{(0),\perp}} \frac{\rmd}{\rmd t} \ket{\Psi^{(0)}}}{2 E} \ket{\Psi^{(0),\perp}} \right) 
    \label{eqn:imag}
\end{aligned}
\end{equation}
On the other hand, equation~(\ref{eqn:B_RealPartofPsi1}) and equation~(\ref{eqn:B_RealPartofPsi2}) give
\begin{multline}
    \epsilon \operatorname{Re} \left( U \ket{\Psi^{(1)}} \right) =  \frac{\cos{\frac{\theta}{2}} \rmi \bra{\Psi^{(0)}} \frac{\rmd}{\rmd t} \ket{\Psi^{(0)}} }{2E + 2 \left. \frac{\rmd h}{\rmd \Sigma} \right\rvert_{\Sigma^{(0)}} \sin^2{\theta}} \begin{pmatrix} 1 & 0 \\ 0 & -\cot{\frac{\theta}{2}} \end{pmatrix} \begin{pmatrix} 1 \\ 1 \end{pmatrix} \;.
\end{multline}
By noticing that $\operatorname{Re}(\rmi \frac{\rmd}{\rmd t} \Psi_1^{(0)}) = 0$, we have  
\begin{multline}
    \label{eqn:B_Miracle}
    \operatorname{Re}\left( \cos{\frac{\theta}{2}} \begin{pmatrix} 1 & 0 \\ 0 & -\cot{\frac{\theta}{2}} \end{pmatrix} \rmi \bra{\Psi^{(0)}} \frac{\rmd}{\rmd t} \ket{\Psi^{(0)}} \begin{pmatrix} 1 \\ 1 \end{pmatrix} \right) = \\ - \operatorname{Re}\left(\rmi U \bra{\Psi^{(0),\perp}} \frac{\rmd}{\rmd t} \ket{\Psi^{(0)}} \ket{\Psi^{(0),\perp}} \right),
\end{multline}
which allows us to write
\begin{multline}
    \epsilon \operatorname{Re} \left( U \ket{\Psi^{(1)}} \right) =  - \operatorname{Re}\left( \rmi U  \frac{\bra{\Psi^{(0),\perp}} \frac{\rmd}{\rmd t} \ket{\Psi^{(0)}}}{2E + 2 \left. \frac{\rmd h}{\rmd \Sigma} \right\rvert_{\Sigma^{(0)}}\sin^2{\theta}} \ket{\Psi^{(0),\perp}} \right)
    \label{eqn:real} \;.
\end{multline}
Finally, combining Eqs.~(\ref{eqn:real}) and (\ref{eqn:imag}) we are able to express the first order term of $\ket{\Psi}$ as 
\begin{multline}
    \label{eqn:B_Psi1}
    \epsilon \ket{\Psi^{(1)}} = - \rmi \frac{\bra{\Psi^{(0),\perp}} \frac{\rmd}{\rmd t} \ket{\Psi^{(0)}}}{2E} \ket{\Psi^{(0),\perp}} \\ + 2 \left. \frac{\rmd h}{\rmd \Sigma} \right\rvert_{\Sigma^{(0)}} \sin^2{\theta} \, U^{\dagger} \operatorname{Re} \left( \rmi U \frac{\bra{\Psi^{(0),\perp}} \frac{\rmd}{\rmd t} \ket{\Psi^{(0)}} \ket{\Psi^{(0),\perp}}}{2E(2E + 2 \left. \frac{\rmd h}{\rmd \Sigma} \right\rvert_{\Sigma^{(0)}} \sin^2{\theta})} \right).
\end{multline}
In the case of the Hamiltonian presented in equation~(\ref{eqn:CompactGPHamiltonian}) in the main text, where $h(\Sigma) = h_z + \frac{g}{2} \Sigma$, we have
\begin{multline}
    \label{eqn:Psi1_ourcase}
    \epsilon \ket{\Psi^{(1)}} = - \rmi \frac{\bra{\Psi^{(0),\perp}} \frac{\rmd}{\rmd t} \ket{\Psi^{(0)}}}{2E} \ket{\Psi^{(0),\perp}} \\ + g \sin^2{\theta} \, U^{\dagger} \operatorname{Re} \left( \rmi U \frac{\bra{\Psi^{(0),\perp}} \frac{\rmd}{\rmd t} \ket{\Psi^{(0)}} \ket{\Psi^{(0),\perp}}}{2E(2E +g \sin^2{\theta})} \right).
\end{multline}
The first term on the right hand side is responsible for the Berry curvature appearing in the expression of the average displacement, while the second term on the right hand side is due to nonlinear effects, and will lead to an additional displacement.

We then move on to the evaluation of the displacement of a particle, whose expression is given by equation~(\ref{eqn:Delta}) in the main text, while taking the average over the state $\ket{\Psi} = \ket{\Psi^{(0)}} + \ket{\Psi^{(1)}}$ with $\ket{\Psi^{(0)}} = \ket{+}$ with eigenenergy $E$.
In equation~(\ref{eqn:Psi1_ourcase}), we rewrite the second term on the right hand side as
\begin{multline}
    \label{eqn:B_Psi1_ourcase_righthand}
     g \sin^2{\theta} \, U^{\dagger} \operatorname{Re} \left( \rmi U \frac{\bra{\Psi^{(0),\perp}} \frac{\rmd}{\rmd t} \ket{\Psi^{(0)}}}{2E(2E -g \sin^2{\theta})} \ket{\Psi^{(0),\perp}} \right) = \\ - \frac{\rmi}{2} \frac{g \sin^2{\theta}}{2E(2E -g \sin^2{\theta})} \left[ \bra{+_{\perp}} \frac{\rmd}{\rmd t} \ket{+} + \bra{+} \frac{\rmd}{\rmd t} \ket{+_{\perp}} \right] \ket{+_{\perp}}.
\end{multline}
Furthermore, we have
\begin{equation}
    \frac{\bra{+_{\perp}} \frac{\partial H}{\partial k} \ket{+}}{2E} = \bra{+_{\perp}} \frac{\partial}{\partial k} \ket{+},
\end{equation} 
\begin{equation}
    -\frac{\bra{+} \frac{\partial H}{\partial k} \ket{+_{\perp}}}{2E} = \bra{+} \frac{\partial}{\partial k} \ket{+_{\perp}}.
\end{equation} 
So we get, up to first order,
\begin{multline}
    \Delta \left< x \right> = \frac{1}{2 \pi} \int \, \rmd t \int_{-\pi}^{\pi} \, \rmd k \, \frac{\partial E}{\partial k} \\ + \rmi \left( \bra{+_{\perp}} \frac{\partial}{\partial t} \ket{+} \bra{+} \frac{\partial}{\partial k} \ket{+_{\perp}} - \bra{+} \frac{\partial}{\partial t} \ket{+_{\perp}} \bra{+_{\perp}} \frac{\partial}{\partial k} \ket{+}  \right) \\
    + \frac{\rmi}{2} \frac{g \sin^2{\theta}}{2E - g \sin^2{\theta}} \left( \bra{+_{\perp}} \frac{\partial}{\partial t} \ket{+} \bra{+} \frac{\partial}{\partial k} \ket{+_{\perp}} \right) \\ - \frac{\rmi}{2} \frac{g \sin^2{\theta}}{2E - g \sin^2{\theta}} \left( \bra{+} \frac{\partial}{\partial t} \ket{+_{\perp}} \bra{+_{\perp}} \frac{\partial}{\partial k} \ket{+}  \right) \\
    + \frac{\rmi}{2} \frac{g \sin^2{\theta}}{2E - g \sin^2{\theta}} \left( \bra{+} \frac{\partial}{\partial t} \ket{+_{\perp}} \bra{+} \frac{\partial}{\partial k} \ket{+_{\perp}} \right) \\ - \frac{\rmi}{2} \frac{g \sin^2{\theta}}{2E - g \sin^2{\theta}} \left( - \bra{+_{\perp}} \frac{\partial}{\partial t} \ket{+} \bra{+_{\perp}} \frac{\partial}{\partial k} \ket{+}  \right).
\end{multline}
The first two terms are also present in the linear case. In particular, we can identify the second term as the Berry curvature $\mathcal{B}(k,t)$, the integral of which yields a quantized Chern number. We can further simplify the third to sixth lines by noticing that
\begin{equation}
    \bra{+} \partial \ket{+_{\perp}} = \partial \left( \frac{\theta}{2} \right) - \rmi \sin{\left( \frac{\theta}{2} \right)} \cos{\left( \frac{\theta}{2} \right)} \partial \phi
\end{equation}
and
\begin{equation}
\begin{aligned}
    \bra{+_{\perp}} \partial \ket{+} &= -\partial \left( \frac{\theta}{2} \right) - \rmi \sin{\left( \frac{\theta}{2} \right)} \cos{\left( \frac{\theta}{2} \right)} \partial \phi \\
    &= - \bra{+} \partial \ket{+_{\perp}} - \rmi \sin{\theta} \partial \phi,
\end{aligned}
\end{equation}
to finally obtain the expression in equation~(\ref{eqn:Displacement}) of the main text.

\section{Apparition of loop structures and singularity points in Bloch state pumping}
\label{app:C}

We consider our GP Hamiltonian 
\begin{equation}
\begin{aligned}
    H(k,t,\ket{\Phi}) &= h_x \sigma_x + h_y \sigma_y + h_z \sigma_z - g\begin{pmatrix} \left|\Phi_{e}\right|^2 & 0 \\ 0 &  \left|\Phi_{o}\right|^2 \end{pmatrix} \\
    H(k,t,\Sigma) &= -\frac{g}{2} I_2 + h_x \sigma_x + h_y \sigma_y + (h_z + \frac{g}{2} \Sigma) \sigma_z,
\end{aligned}
\label{eqn:C_Hgeneral}
\end{equation}
where $\Sigma = \abs{\Phi_o}^2 - \abs{\Phi_e}^2$ and $h_x, h_y, h_z$ are the same coefficients used in equation~(\ref{eqn:GPHamiltonian}) of the main text. Based on the known eigenstate solutions for two-level systems, we can find a stationary state of the form
\begin{equation}
    \ket{\Phi} = \begin{pmatrix} \cos{\frac{\theta}{2}} \\ \sin{\frac{\theta}{2}} e^{\rmi \phi} \end{pmatrix} = \frac{1}{\sqrt{2}} \begin{pmatrix} \sqrt{1-\Sigma} \\ \sqrt{1+\Sigma} e^{\rmi \phi} \end{pmatrix}
    \label{eqn:C_Northgauge}
\end{equation}
where $e^{\rmi \phi} = \frac{h_x + \rmi h_y}{\sqrt{h_x^2 + h_y^2}}$, and $-1 < \Sigma < 1$ which satisfies the self-consistency equation
\begin{equation}
    \begin{aligned}
    \cos{\theta} &= \frac{h_z + \frac{g}{2}\Sigma}{\sqrt{h_x^2 + h_y^2 + (h_z + \frac{g}{2}\Sigma)^2}} \\
    - \Sigma &= \frac{h_z + \frac{g}{2}\Sigma}{\sqrt{h_x^2 + h_y^2 + (h_z + \frac{g}{2}\Sigma)^2}} \\
    0 &= [h_x^2 + h_y^2 + (h_z + \frac{g}{2}\Sigma)^2]\Sigma^2 - (h_z + \frac{g}{2}\Sigma)^2.
    \end{aligned}
    \label{eqn:C_SelfConsistency}
\end{equation}
As a quartic in $\Sigma$, this self-consistency equation can have either 2 or 4 real solutions, the latter case implying the existence of additional bands, e.g., loop structures. 

In our case, loop structures first appear at $\omega t = \frac{\pi}{2},\frac{3 \pi}{2}$. In this case, we have $h_z = 0$ and the self-consistency equation can be factorized as 
\begin{equation}
     \Sigma^2 \left( \Sigma^2 - \frac{g^2-4(h_x^2 + h_y^2)}{g^2} \right) = 0,
     \label{eqn:C_SelfConsistencySpecialPoint}
\end{equation}
which always has a double root $\Sigma = 0$ with energies $E =\pm \sqrt{h_x^2 + h_y^2}$. However, if the nonlinearity is strong enough such that $2\sqrt{h_x^2 + h_y^2} \leq \abs{g}$ there are two additional solutions, $\Sigma = \pm \sqrt{\frac{g^2 - 4(h_x^2 + h_y^2)}{g^2}}$, both with energy $E=-\frac{g}{2}$.
In our case, $h_x^2 + h_y^2 = 2[J^2 + \delta^2 \sin^2{(\omega t)}] + 2 \cos{k} [J^2 - \delta^2 \sin^2{(\omega t)}]$. So for $\omega t = \frac{\pi}{2},\frac{3 \pi}{2}$, we have $h_x^2 + h_y^2 = 2[J^2 + \delta^2] + 2 \cos{k} [J^2 - \delta^2]$ which is minimal for $k=\pi$, giving $h_x^2 + h_y^2 = 4 \delta^2$
Hence, the lowest value of $g$ for which we have loop structures is $g_c = 4 \delta = 2$.

Using $\Sigma = -\cos{\theta}$ and $\partial \Sigma = \sin{\theta} \, \partial \theta$ (where $\partial$ can be any partial derivative, $\partial_t$ or $\partial_k$), we can rewrite $\mathcal{D}(k,t)$ in equation~(\ref{eqn:Drift}) of the main text as
\begin{equation}
    \mathcal{D}(k,t) = -\frac{1}{2} \frac{g (1 - \Sigma^2)}{2E + g (1 - \Sigma^2)} \partial_t \phi \, \partial_k \Sigma.
    \label{eqn:C_DriftSimple}
\end{equation}
The Berry curvature can also be simplified in a similar fashion. First, we express the Berry connection as 
\begin{equation}
\begin{aligned}
        \mathcal{A}_{\alpha}  &= \rmi \braket{\Phi}{\partial_{\alpha}\Phi} \\
        &= -  \frac{1 + \Sigma}{2} \partial_{\alpha} \phi,
\end{aligned}
\end{equation}
where $\alpha$ can be $t$ or $k$. Then, using $\mathcal{B}(k,t) = \partial_t \mathcal{A}_k - \partial_k \mathcal{A}_t$ we get
\begin{equation}
    \mathcal{B}(k,t) = \frac{1}{2} (\partial_t \phi \, \partial_k \Sigma - \partial_k \phi \, \partial_t \Sigma).
    \label{eqn:C_BerryCurvatureSimple}
\end{equation}
At points $(k,t) = (\pi,\frac{\pi}{2})$ and $(k,t) = (\pi,\frac{3 \pi}{2})$  using the self-consistency equation given in equation~(\ref{eqn:C_SelfConsistencySpecialPoint}), we also get
\begin{equation}
    \partial \Sigma = \frac{[g^2 - 8(h_x \partial h_x + h_y \partial h_y)] \Sigma }{4 g^2 \Sigma^2 - 2[g^2 - 4(h_x^2 + h_y^2)]}.
    \label{eqn:C_PartialSigma}
\end{equation}
By considering the solution $\Sigma = \sqrt{\frac{g^2 - 4(h_x^2 + h_y^2)}{g^2}}$ in equations (\ref{eqn:C_DriftSimple}), (\ref{eqn:C_BerryCurvatureSimple}) and (\ref{eqn:C_PartialSigma}), we get
\begin{equation}
\begin{aligned}
    \mathcal{D}(k,t) &= \frac{g}{2} \frac{8 \delta^2 \partial_t \phi}{(g^2 - 16 \delta^2)^{\frac{3}{2}}} \\
    \mathcal{B}(k,t) &= \frac{g}{4} \frac{\partial_t \phi - \partial_k \phi}{(g^2 - 16 \delta^2)^{\frac{1}{2}}}.
\end{aligned}
\end{equation}
where $\partial_t \phi$ and $\partial_k \phi$ do not depend on $g$. We can then see that both quantities exhibit a singularity point at the critical value $g_c = 4 \delta$.

\section*{References}
\bibliographystyle{iopart-num}
\bibliography{Bibliography}

\end{document}